\def\seq{{\rm sin}^2 \vartheta}
\def\coq{{\rm cos}^2 \vartheta}
\def\be{\begin{equation}}
\def\ee{\end{equation}}
\def\bea{\begin{eqnarray}}
\def\eea{\end{eqnarray}}

\def\ka{\kappa_{0}}

\def\va{\vartheta}
\def\arqo{\frac{a^2}{r_0^2}}
\def\M{\cal M}

\def\S{\cal S}

\def\A{\cal A}
\def\Seq{{\rm sin}^2 \vartheta_{+}}

\documentstyle[epsf]{article}
\begin{document}
\onecolumn
\begin{flushright}
Alberta-Thy-16-94 \\
%gr-qc/xxxx
\end{flushright}
\vfill
\begin{center}
{\Large \bf Mass Inflation in a Rotating Charged
Black Hole}\\
\vfill
Alfio Bonanno\\
%\footnote[3]{Permanet address: Institute of Astronomy, University of
%Catania,
%Viale Andrea Doria 5, 95125 Catania, Italy}, S. Droz, W. Israel and S.M.
%Morsink\\
\vspace{2 cm}
{\em Canadian Institute for Advanced Research Cosmology Program, \\
Theoretical Physics Institute, University of  Alberta,\\
Edmonton, Alberta, Canada T6G 2J1\\
and\\
Institute of Astronomy, University of Catania\\
Viale Andrea Doria 6, 95125 Catania, Italy}\\
\vspace{2 cm}
\vfill
%%%%%%%%%%%%%%%%%%%%%%%%%%%%%%%%%%%%%%%%%%%%%%%%
%
%ABSTRACT
%
%%%%%%%%%%%%%%%%%%%%%%%%%%%%%%%%%%%%%%%%%%%%%%%%
\begin{abstract}
The structure of the Cauchy horizon of a charged rotating black hole
is analyzed under the combined effect of an ingoing and
outgoing flux  of gravitational waves.  
In particular, by means of  
an axisymmetric realization of the Ori model, the growth of the mass 
parameter near the Cauchy horizon is studied in the slow rotation 
approximation. It is shown that the mass-parameter inflates,
while the angular momentum per unit mass deflates, but
initial deviations from spherical symmetry survive.
%The asymptotic configuration of the 
%spacetime near the Cauchy horizon is characterized by a
%``mass-inflated'' Kerr geometry where deviations from the
%spherical geometry in the Cauchy horizon reflects
\end{abstract}
\vfill
\end{center}
%\clearpage
%%%%%%%%%%%%%%%%%%%%%%%%%%%%%%%%%%%%%%%%%%%%%%%%
%
%
%
%%%%%%%%%%%%%%%%%%%%%%%%%%%%%%%%%%%%%%%%%%%%%%%%

\twocolumn

Although it is now generally accepted that gravitational 
collapse results in the formation of a black hole,  the ultimate 
fate of the collapsing object within the black hole is an open 
question. The presence 
of an inner horizon, - the Cauchy horizon
(CH) - a lightlike surface behind which the predictability of the 
field equations breaks down, turns out to
be a formidable obstacle to constructing an unambiguous picture of the 
complete analytical extension of the geometry. 
This fundamental issue is encoded in the 
peculiar character of the CH. 

As noted first by Penrose \cite{penro}, 
ingoing pencils of radiation 
experience a diverging blueshift
as they approach the generators of the CH.  This kind of nonscalar
singularity, known as  a ``whimper'' \cite{ek}, is unstable to
perturbations. Then a stronger, scalar singularity can develop when 
the backreaction of the fields on the metric is taken into account.
If the additional effect of an outgoing flux is considered, 
spherical models of the crossflow region \cite{pi} show that 
the effective mass parameter $m(u,v)$ exponentially inflates 
at late advanced times, as the CH is approached. 
In particular the Weyl curvature 
invariant $\Psi_{2}$ diverges, indicating
that a scalar singularity occurs. But this divergence is 
still  ``mild", since the mass function is an integrable 
function of the Kruskalized advanced coordinate, and 
in suitable coordinates the metric coefficients 
stay finite at the CH.

How general are these models in describing the 
evolution of the interior at late advanced times, if the restriction of 
spherical symmetry is removed?  
General arguments based on the constancy of the surface gravity over
a stationary horizon \cite{pi} indicate that  the growth of the mass 
parameter should appear uniform on small angular scales.
One suspects in particular that
in a generic axisymmetric collapse the ``effective Kerr parameter'' 
(the angular momentum per unit mass $a=J/m$ ) becomes negligible 
if the total angular momentum of the
inflalling radiation is bounded during the collape 
\cite{pbi}. The asymptotic structure of 
the spacetime close to the
CH should look like an axially symmetric  
geometry with an enormously inflated mass term.
Examples of axisymmetric mass-inflation solutions have been discussed 
in  $2+1$ models
\cite{ma}, however the resulting Kretschmann invariant is
found to be finite at the singularity unlike the
spherically symmetric model in $3+1$ dimensions. 
A more realistic analysis 
of the instability of the CH  for a class of Kerr-Newman  
spacetime has been proposed in \cite{dk}, and in  the framework of the 
$2+2$ approach in \cite{pbm}. In this latter analysis the
resulting asymptotic configuration
seems not to be that of a Petrov type D spacetime. 

A possible insight into the question lies in the nature of the 
mass-inflation phenomenon. The outgoing 
flux is a catalyzer which causes the generators
to contract without a direct interaction
with  the infinitely blueshifted 
infalling lightlike contribution.  The rate of contraction 
is fully determined by Price's
power law damping of the radiative tail $\sim 1/v^{(p-1)}$, $p\geq 11$. 
Hence one can argue that deviations from the purely spherical
geometry of the CH should be reflected in deviation from 
spherical symmetry in the mass-inflated sector, since the contraction 
will not be uniform in a non-spherical model. 
The leading contribution to the mass function should 
then be dominated by a very large mass term with a small
angular dependence.
We shall here present an explicit mass-inflation solution in 
the case of a rotating charged hole that exhibits this behavior. 
Our model, although approximate, should capture the qualitative
features of the geometry of the CH in
a non-stationary, rotating black hole.

The crossflow region, near the CH, is 
described by an outgoing lightlike shell - simulating the outgoing
flux - embedded in a continuous flow of infalling gravitational waves,
see Fig.1 . 
This axisymmetric realization of the Ori 
\cite{ori} model
is derived in the slow rotation approximation when $a=\frac{J}{m}$
is small compared to the radius of the CH $r_0$ 
\be
\epsilon\equiv\arqo\ll1 \label{prima}
\ee
In particular, as background geometry we consider the 
non-stationary Vaidya type  
generalization of the Kerr metric discussed in \cite{ck}. We 
extend that model to the charged case in order to retain a finite
(nonzero) radius for the inner horizon even when $a$ is small.
In the $\{r,\vartheta,\varphi,v\}$ 
Eddington-Kerr coordinate system 
the metric reads
\bea
&&ds^2=-\left (1-\frac{2m\left (v\right )r-e^2}{\Sigma}\right )
dv^2+2drdv\nonumber\\[2 mm]
&&+\Sigma 
d\vartheta^2-2a\seq d\varphi dr
+{\cal R}^2\seq d\varphi^2 \nonumber\\[2 mm]
&&-\frac{2a(2m\left (v\right )r-e^2)
\seq}{\Sigma}dv d\varphi\label{met}
\eea
where
\bea
&&{\cal R}^2=
\frac{(r^2+a^2)^2-\Delta a^2\seq 
}{\Sigma},\nonumber\\[2 mm]
&&\Sigma=r^2+a^2\coq,\nonumber\\[2 mm]
&&\Delta
= r^2+a^2+e^2-2m(v)r
\eea 
As in the spherical Vaidya model, the mass 
parameter $m(v)$ is a function of the advanced coordinate $v$.  Its
functional dependence near the CH, located at $v=+\infty$, is assumed 
to be of the form 
\be
m=m_{0}+\delta m(v),\quad\delta m(v)=-\frac{A}{v^{p-1}}\label{deppa}
\ee
to reproduce the power law decay of the gravitational waves.
Although in a realistic setting the mass function has probably an 
angular dependence as well, the assumption (\ref{deppa})
is physically reasonable 
since we work in the slow rotation approximation, and 
(\ref{deppa}) is the only relevant contribution.  
In the following analysis it is convenient 
to introduce the null complex 
tetrad $\{ l_{\mu},n_{\nu},
m_{\mu},\bar{m}_{\mu} \}$, with $-l_{\mu}n^{\mu}=m_{\mu}{\bar{m}}^{\mu}
=1$ and $l_\mu$ is the 
repeated principal null direction associated with
the infalling field - $r$ decreases with time 
along $v={\rm const.}$ - 
\bea
&&l_\mu=-\partial_{\mu}v+a{\rm sin}^2\vartheta
\partial_{\mu}\varphi\nonumber\\[2 mm]
&&n_{\mu}=-\frac{1}{\Sigma}(\frac{\Delta}{2}
\partial_{\mu} v-\Sigma\partial_\mu r-a
{\rm sin}^2\vartheta \frac{\Delta}{2}\partial_\mu \varphi)
\nonumber\\[2 mm]
&&m_\mu=-\frac{\bar{\rho}}{\sqrt{2}}(i a {\rm sin}\vartheta
\partial_\mu v+\Sigma\partial_\mu\vartheta\nonumber\\[2 mm]
&&-i\left (r^2+a^2\right ){\rm 
sin} \vartheta\partial_\mu\varphi )\label{tetra}
\eea
where $\rho=-(r-i a {\rm cos}\vartheta)^{-1}$. 
The total energy momentum tensor can be expressed as
\bea
&&T_{\mu\nu}=2\phi_{22}l_\mu l_\nu-4\phi_{12}l_{(\mu}\bar{m}_{\nu)}-
4\bar{\phi}_{12}
l_{(\mu}m_{\nu)}\nonumber\\[2 mm]
&&+4\phi_{11}(l_{(\mu}n_{\nu)}+m_{(\mu}\bar{m}_{\nu)})\label{stre}
\eea
where $\phi_{22}, \phi_{12}, \phi_{11}$ are the only non-vanishing tetrad
components of the trace free part of the Ricci tensor 
$S_{\mu\nu}=R_{\mu\nu}-1/4g_{\mu\nu}R,$ ($=R_{\mu\nu}$ in our case),
\bea
&&\phi_{22}=r(2\dot{m}(v)-a^2\seq \ddot{m}(v))/4
\Sigma^2\nonumber\\[2 mm]
&&\phi_{11}=e^2/2\Sigma^2,\nonumber\\[2 mm]
&&\phi_{12}=-ia{\rm sin}\vartheta\dot{m}(v)\rho/2
\sqrt{2}\Sigma\label{rite}
\eea
The radiation field consists of a pure null part and a residual term.
In particular $\phi_{11}$ represents the contribution of a static source 
field generated by a charge of strength $e$.
Unlike the stationary case, there
is only one repeated principal null direction:
the spacetime is algebraically 
special of Petrov type II and the only 
non-zero Weyl invariants are
\bea
&&\Psi_{2}=-m(v)\rho^3 -e^2\rho\bar{\rho}^3\nonumber\\[2 mm]
&&\Psi_{3}=\frac{-i\dot{m}(v)\rho
a {\rm sin}\vartheta}{2\sqrt{2}\Sigma}
-\frac{i\dot{m}(v) \rho^2ra {\rm sin}\vartheta }{\sqrt{2}\Sigma}
\nonumber\\[2 mm]
&&\Psi_{4}=\frac{\ddot{m}(v) \rho^2 ra^2 \seq }{\sqrt{2}\Sigma}
+\frac{\dot{m}(v) \rho^3 ra^2 \seq }{\sqrt{2}\Sigma}\label{psi}
\eea

Let us now consider the equation of motion 
$r=r(v,\vartheta)$ of an axisymmetric generic 
outgoing null hypersurface. Close to CH, 
where the functional dependence of the
mass function on the advanced coordinate is given by (\ref{deppa}), 
it reads
\bea
&&-\frac{2\delta m(v)r}{r^2+a^2}-2\ka(r-r_0) - 2
\partial_{v}r\nonumber\\[2 mm]
&&+\frac{a^2\seq}{r^2+a^2}(\partial_{v}r)^2
+\frac{(\partial_{\vartheta}r)^2}{r^2+a^2}=0\label{moto}
\eea
where $\ka$ is the surface of gravity of the inner horizon
located at $r=r_0$.
To first order in the 
effective rotation parameter $\epsilon$,
the solution of has to be of the form
\be
r-r_0=f(v)+\epsilon g(v,\va)\label{ansa}+O(\epsilon^2)\label{erre}
\ee
and (\ref{moto}) is equivalent to the following 
system of partial differential equations for the zeroth and 
first order terms respectively
\bea
&&\ka f+f_{v}=-
\frac{\delta m(v)}{r}\nonumber\\[2 mm]
&&\ka g+g_v-\frac{1}{2}\seq f_v^2=\frac{\delta m(v)}{r}\label{siste}
\eea
The boundary conditions 
\be
\lim_{v\rightarrow\infty} f(v)=g(v,\va)=0\label{bcs}
\ee
determine the following asymptotic form for the solution as 
$v\rightarrow\infty$:
\be
r-r_0=f(v)\left (1-\epsilon\left (\frac{\seq (p-1)}{2\ka v}+
O(\epsilon/v^2)\right)\right )\label{sol2}
\ee
where 
\be
f(v)=\frac{A}{\ka r_0}v^{-(p-1)}(1+\frac{p-1}{\ka v}+\dots)\label{sol3}
\ee
This shows that the angular dependence is suppressed by a factor $1/v$ 
to first order in $\epsilon$ as we approach the CH.   

Now we consider this hypersurface to be the locus
of a  lightlike shell embedded in this background.
The spacetime is then divided in two regions 
$\M^{+}$,and $\M^{-}$, see Fig.1, separated by the 
outgoing shell $\S$ whose equation of  motion is
of the form (\ref{sol2}) near the CH. We assume that 
the ``past'' side of the shell is described by the 
radiating Kerr-Newman metric (\ref{met}). 
In general nothing can be said  about the future side of $\cal S$. 
However, it is reasonable to think that in the slow rotation regime
the production of gravitational waves can be neglected and 
the structure of the resulting ``glued''  manifold
is still Kerr-like at least close to CH.
Thus as ``trial'' metric we assume that
the spacetime in the future sector of the shell
can be represented with
a line element of the type (\ref{met}) where the coefficients
functions $r_{+}, a_{+}, m_{+}$
depend on the coordinates in  $\M^{-}$.
\vspace*{0.5cm}\\
\epsfxsize=7cm
\epsffile{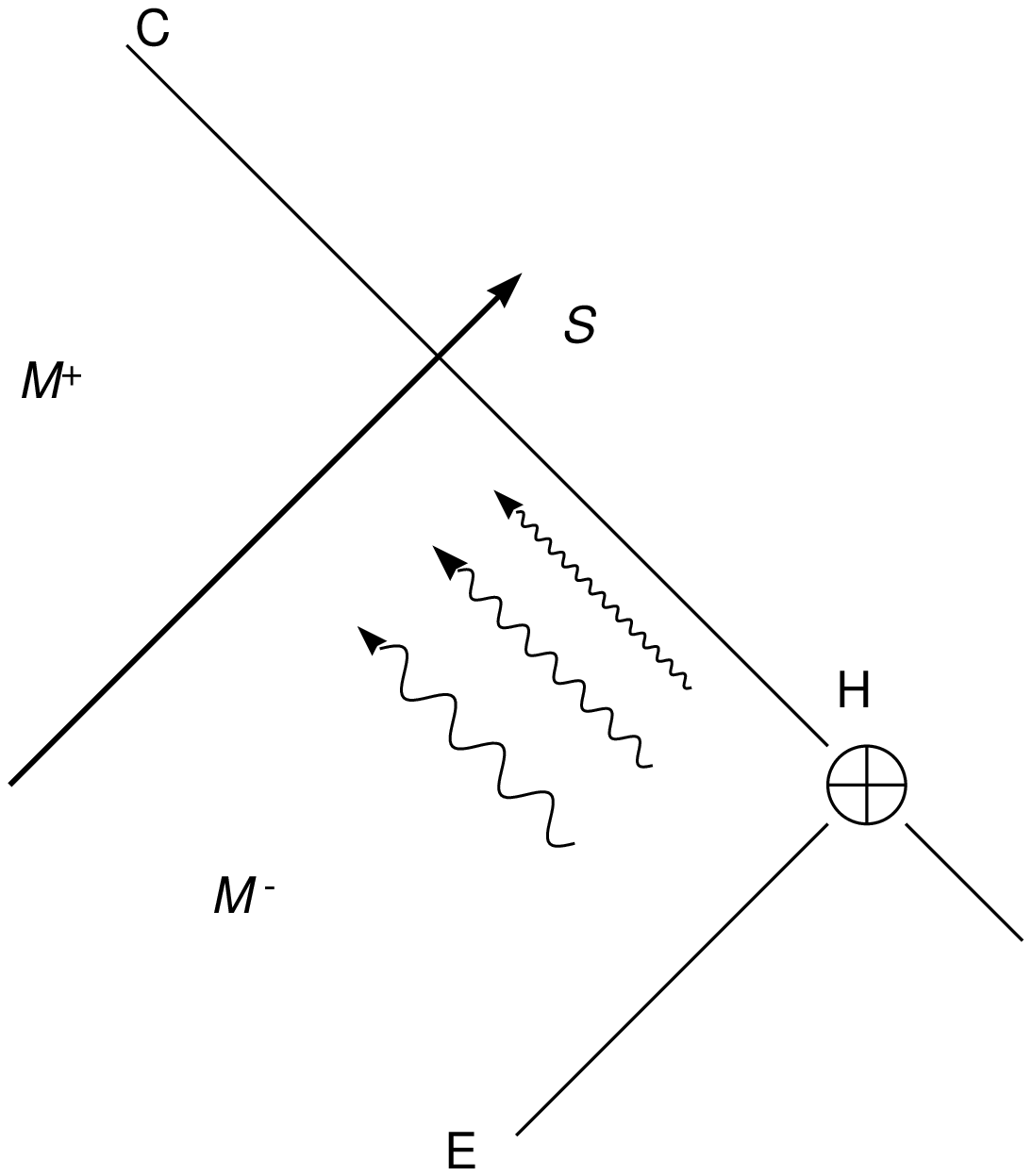}
\begin{center}
\parbox{6.5cm}{
%\baselineskip 13pt
%\vspace*{0.5cm}
\small
Fig. 1. A spacetime diagram of the
equatorial plane of the axisymmetric
black hole interior showing
the ingoing flow of gravitational waves being
crossed by an outgoing lightlike
shell infalling in the Cauchy horizon.}
\end{center}
\mbox{\hspace{1cm}}
\vspace*{0.1cm}\\
In particular let $\{v_{\pm}$,$r_{\pm}$,$\vartheta_{\pm}\}$
be the local coordinates
of $\S$ in $\M^{\pm}$, $a_{\pm}$
the value of the angular momentum per unit mass, and 
 $m_{\pm}$ the mass functions in both the sides of the shell.
The stress-energy tensor in $\M^{+}$ contains
the contribution in (\ref{stre}) and residuals
terms arising from the fact that $a_{+}$ is not stricly
constant. Those terms can be shown to be much
smaller than the leading, optical geometric
contribution in (\ref{stre}) close to CH.
We shall explicitly check the
validity of this approximation
at the end of the computation.

In this model the presence of  the outgoing 
lightlike shell simply serves to start the contraction of 
the generators of the CH.  
Thus we consider a pressureless shell so that the soldering 
of the two geometries is affinely conciliable \cite{wb}. 
We remark that since  
the analysis in \cite{ab,ab1} shows that for spherical 
symmetry the CH survives the focusing 
effect of the outgoing flux, it is reasonable to think that this 
would be the case even if the hole is slowly rotating. 
In order to isolate the divergent contribution in the 
mass function in $\M^{+}$ we define $m_{+}=\bar{m}+M(v_{+})$
where, by definition $2\bar{m}\bar{r}=\bar{r}^2+\bar{a}^2+e^2$ and
$\bar{r}$, $\bar{a}$ are the values of $r_{+}$ and $a_{+}$ at the CH. As
before, the solution of  (\ref{moto}) in $\M^{+}$  is of the form
\be
r_{+}-r_{0+} =f _{+}(v_{+})+\frac{a_{+}^2}{r^2}g_{+}
(v_{+},\vartheta_{+})\label{rana}
\ee
Therefore equation (\ref{rana}) decouples as a 
system of the type (\ref{siste}). In particular if  
$M(v_{+})\gg r_0$ near the CH, at the 
leading order in the mass term it reduces to
\be
\frac{\partial f_{+}}{\partial v_{+}} \simeq-
\frac{M_{+}}{r_{0+}},\quad\quad
\frac{\partial g_{+}}{\partial v_{+}}
\simeq\Seq\frac{M^2(v_{+})}{r_{0+}^2}
\label{lld}
\ee

The only geometric condition that has to be satisfied 
along $\S$, the common boundary of the two spacetimes, is that
the two intrinsic degenerate metrics coincide. This implies 
that the area ${\A}$ of the two intrinsic metrics has to be continuous
across the shell,
\be
[{\A}]=0 \label{conti}
\ee
where $[{\A}]=\A_{+}-\A_{-}$. In a perturbative expansion in $a_\pm$, 
this condition decouples into  two
distinct continuity requirements for the zeroth and first order terms.
The spherically symmetric contribution simply states the 
continuity of the $r$ coordinate across $\S$, not {\em a priori}
guaranteed from
(\ref{conti}). Thus $r_{+}=r_{-}=r$ and we set $r_{0+}=r_{0-}=r_0$ 
at the CH.  We use $r$ as a parameter 
(necessarily affine for a pressureless shell) 
along the generators of $\S$
By using eq. (\ref{sol2}) 
in $\M^{-}$ one finds that along the shell 
the area of any 
$v_{-}={\rm const.}$ hypersurface reads 
\be
{\A}_{-}=4\pi r_{0}^2(1+\epsilon)+O(1/v^{p}), 
\quad\quad v_{-}\rightarrow +\infty\label{auno}
\ee 
where only linear term in $\epsilon$ have 
been retained in the degenerate metric.        
Similarly in $\M^{+}$, close to the CH and 
up to linear terms to $a_{+}^2$, one has
\be
{\A_{+}}=4\pi\left (r_{0}^2+a_{+}^2\left 
(1+\frac{2M(v_{+})}{3r_{0}}\right )\right )\label{adue}
\ee
In particular, if $M(v_{+})/r_0\gg 1$, equation (\ref{conti})
reads
\be
a^2_{+}M_{+}\simeq 3\epsilon r_{0}^3/2, 
\quad v_{-}\rightarrow +\infty\label{fund}
\ee 
This latter equation contains the essential physics. 
As observed in the beginning, it explicitly 
shows that the effective Kerr parameter $a_{+}$ becomes 
increasingly small as the mass function grows. 
We stress, however, that the asymptotic geometry close to the CH 
is not spherically symmetric because 
the $a^2_{+}M_{+}$ and $a_{+}M_{+}$ terms 
in the metric are not negligible. 
The dependence of $v_{+}$ and $\vartheta_{+}$ on the 
advanced coordinates in the past sector of the shell can be
determined from the expression for the null generators
on both sides of $\S$ (the dependence of  
$\varphi_{+}$ on $\varphi_{-}$ is trivial since these coordinate
define the same Killing vector). 
{}From the chain rule and the continuity of the $r$ 
function one has
\bea
&&\partial_{v}f_{-}=\partial_{v}f_{+}\left [\frac{\partial v_{+}}
{\partial v_{-}}\right ]^{(0)}
\nonumber\\[2 mm]
&&\epsilon \partial_{v}g_{-}=\partial_{v}f_{+}
\left [\frac{\partial v_{+}}{\partial v_{-}}\right ]^{(1)}
+\frac{a_{+}^2}{r_{0}^2}\partial_{v}g_{+}
\left [\frac{\partial v_{+}}{\partial v_{-}}\right ]^{(0)}
\nonumber\\[2 mm]
&&\epsilon \partial_{\vartheta}g_{-}=\partial_{v}f_{+}
\left [\frac{\partial v_{+}}{\partial \vartheta_{-}}\right ]^{(1)}
+\frac{a_{+}^2}{r_{0}^2}\partial_{\vartheta}g_{+}
\label{m3}
\eea
where only first order terms in the 
jacobian determinant have been retained. 
To relate the dynamics of the two spacetime one has to
add the condition for matching of normal
stresses across the shell
\be
[T^{\mu\nu}s_{\mu}s_{\nu}]=0\label{nma}
\ee
where $s_{\mu}$ are generators of $\S$.
Equation (\ref{nma}) is a second order ordinary differential equation
for $M_{+}$. The dependence of  $\{v_{+},\vartheta_{+}\}$ on the 
local coordinates in $\M_{-}$ is implicitly 
defined from (\ref{m3}), and from the
equations of motion of $\S$ in the two spacetimes.  
The explicit solution of (\ref{nma}) is not
available.  However in the slow rotation approximation 
any scalar product between the generators $s^{\mu}$ and the 
tetrad vector must be written as a sum of a function of only 
the advanced coordinated plus $\epsilon$ times a function
of both advanced and angular coordinates,  and
(\ref{nma}) reduces to 
\be
[\phi_{22}l_{\mu}l_{\nu}s^{\mu}s^{\nu}]=0\label{fite}
\ee
where the scalar products have to be calculated from the 
spherically symmetric contribution. Therefore
it is important to stress that in this approximation
only the optical, physically meaningful,
part of the energy momentum tensor is relevant.  
This equation contains the coupling between the 
angular momentum and mass function,
through the $a_{+}\ddot{M}_{+}$ 
term.  By using 
(\ref{m3}) in (\ref{fite}) and by 
expressing $a_{+}$ with the help of eq. (\ref{fund}), 
after some simplifications one explicitly finds
\bea
&&\left [\frac{d^2M_{+}}{dv_{-}^2}+
\frac{1}{M_{+}}\left (\frac{dM_{+}}{dv_{-}}\right )^2\right ]
\frac{3\epsilon r_{0}\seq }{4\partial_{v}f_{-}}-
\frac{dM_{+}}{dv_{-}}\nonumber\\[2 mm]
&&=-\ka\left (1-
p\frac{4+\ka r_03\epsilon\seq}{4\ka v_{-}}\right ) M_{+}\label{eqm}
\eea
note that $\vartheta_{-}=\vartheta_{+}$ in this approximation. 
Hence we write
\be
M_{+}=m_{+}+\delta m_{+}
\ee
and we have from (\ref{eqm})
\be
m_{+}=\frac{1}{v^p_{-}}e^{\ka v_{-}},
\quad v_{-}\rightarrow+\infty\label{ipo}.
\ee
This has the Israel-Poisson behavior, and
\be
\delta m_{+}\sim c\epsilon\seq e^{\ka v_{-}}
\ee
where $c$ is a constant.
This shows  the effective mass parameter exponentially 
inflates with a residual angular dependence 
\be
M_{+}\sim e^{\ka v_{-}}(1+c\epsilon\seq)\label{massa}
\ee
to first order in $\epsilon$, that does not effect the
exponentially divergent prefactor \cite{pi}.
{}From the equations (\ref{m3}) one finds
\bea
&&\left [\frac{\partial v_{+}}
{\partial v_{-}}\right ]^{(0)}\sim e^{-\ka v_{-}},\quad
\left [\frac{\partial v_{+}}
{\partial v_{-}}\right ]^{(1)}
\sim \epsilon {\rm sin}^2\vartheta_{-}e^{-\ka v_{-}}\nonumber\\[2 mm]
&&\left [\frac{\partial v_{+}}
{\partial \vartheta_{-}}\right ]^{(1)}\sim 
\epsilon\ka^{-1}{\rm sin}2\vartheta e^{-\ka v_{-}}
\eea
We see that the radial coordinate 
tends to a finite limit {\em behind} the shell. Indeed, 
from eq. (\ref{lld}), we have
\bea
&&f(v_{+})\sim\frac{1}{\ln^{p}|v_{+}|},\quad\quad
v_{+}\rightarrow 0\nonumber\\[2 mm]
&&a^2_{+}g(v_{+},\vartheta_{+})\sim\frac{r_{0}^2\Seq}
{\ln^{p}|v_{+}|},\quad\quad v_{+}\rightarrow 0
\eea
thus, 
\be
\lim_{v_{+}\rightarrow 0}{r(v_{+})}=r_{0}
\ee
The geometry in the mass-inflated sector is asymptotically dominated 
by the large mass term, and the metric, to a good approximation,
explicitly  reads 
\bea
&&ds_{+}^2\approx\frac{2m_{+}}{r}(1+c\epsilon\seq)dv_{+}^2+2drdv_{+}
+r^2d\vartheta^2\nonumber\\[2 mm]
&&+r^2(1+3\epsilon\seq)\seq d\varphi^2\nonumber\\[2 mm]
&&-4\sqrt{3\epsilon m_{+}r/2}
\seq dv_{+}d\varphi\label{me1}
\eea
The following ``mild'' twist  of the $\varphi$ coordinate
(since $m_{+}$ is an integrable function of the 
advanced coordinate $v_{+}$)
\be
d\varphi=d\Phi+\frac{2\sqrt{3\epsilon m_{+}r_{0}/2}}{r_{0}^2}dv_{+}
\ee
brings the metric in the final form
\bea
&&ds_{+}^2\approx\frac{2m_{+}}{r}(1+c\epsilon\seq)dv_{+}^2+
2drdv_{+}\nonumber\\[2 mm]
&&+r^2d\vartheta^2+
r^2(1+3\epsilon\seq)\seq d\Phi^2\label{me2}
\eea
to linear terms in $\epsilon$. This result is interesting, 
we believe, because it explicitly shows that deviations
from spherical symmetry at the Cauchy horizon
are reflected in the mass-inflated sector,
in accordance with the remarks
at the beginning. This phenomenon
should be characteristic of the dynamics
of a spacetime with a non-spherical CH. 
It is important to check the
consistency of the approximations that have been done. In fact
had we {\em started} with the line element (\ref{me2}),
with $m_{+}=v_{+}^{-1}{\rm ln}^{-p}|v_{+}|$  we would find 
\be
{\A}_{+}=4\pi r_{0}^2(1+\epsilon)={\A}_{-}+O(1/v^{p})
\ee   
for the continuity of the intrinsic area across the shell.
Also, it is straightforward to verify that 
the stress-energy tensor computed from (\ref{me2}) close to
CH satisfies the matching condition for 
the normal stresses, equation (\ref{nma})
\be
[T^{\mu\nu}s_{\mu}s_{\nu}]=O(\epsilon^2).\label{nma1}
\ee
The Komar invariant quantity associated with the rotational Killing 
vecor field $\xi^{\mu}$ is not conserved 
since matter is flowing into the system. 
It is indeed divergent in our model, but with a much slower rate
\be
\frac{1}{8\pi}\oint \xi^{\mu;\nu}
d\sigma_{\mu\nu}\sim\sqrt{m_{+}}\label{ang}
\ee
where the integral is taken over the two dimensional boundary of
any $v_{-}={\rm const.}$ hypersurface, with $v_{-}\rightarrow+\infty$.
It is however hard to judge, from the analysis here presented, whether 
one can expect this latter result to occur in a more general 
framework than that of our model.
In particular at the present we do not see
any deeper physical argument to explain it.

As in the sperically symmetric models, at the CH 
a strong, scalar singularity develops, whose character 
can be read off from the Weyl 
curvature invariants in (\ref{psi}). One finds that to the future of  
the shell, they are all divergent:
\bea
&&\Psi_{2}\sim \frac{1}
{v_{+}\ln^{p} |v_{+}|},
\quad\Psi_{3}\sim \frac{1}{v_{+}\sqrt{|v_{+}|}\ln^{p/2} |v_{+}|}
\nonumber\\[2 mm]
&&\Psi_{4}\sim \frac{1}{v^2_{+}\sqrt{|v_{+}|}\ln^{p/2} |v_{+}|}
\eea
Although $\Psi_{3}$ and $\Psi_{4}$ are tetrad-dependent,
the divergence of the boost-invariant  quantity $\Psi_2$  
has the same ``mild''  trait as in the spherically symmetric case. 

Finally, we remark that these results  have been derived in the
slow rotation approximation and it would be interesting 
to see how this scenario would evolve  in the more 
general case of arbitrary spin.
%\acknowledgments
\vspace{1 cm}

The author would like to thank Werner Israel for his constant advice
and many enlightening discussions, as well Valery Frolov, 
Sharon Morsink, Charles Torre,
David Hobill, Roberto Balbinot
and Eric Poisson, for many 
useful and important comments.   
This work has been supported by the Italian Minister for
the University and Scientific Research,and by the
Institute of Astronomy of the University of Catania.


\begin{thebibliography}{12}
\bibitem{penro}R.Penrose in {\it Battelle Rencontres}, ed. 
C.M.DeWitt and J.A.Wheeler, W.A.Benjamin, New York, 1968,
p.222
\bibitem{ek}G.F.R. Ellis and A.R. 
King, {\em Commun. Math. Phys.} {\bf 38}, 119
(1974). A.R. King, {\em Phys. Rev.} {\bf D 11}, 763 (1975).
\bibitem{pi}E. Poisson and W. Israel, {\em Phys. Rev.}
{\bf D41}, 1796, (1990).
\bibitem{pb}P.R.Brady and J.D. Smith, 
``Black hole singularities: a numerical approach"
Univ. of Newcastle-Upon-Tyne, to appear in 
{\em Physical Review Letters}. 
\bibitem{pbi}C.Barrabes, W. Israel, 
E. Poisson, {\em Class. Quantum Grav.}, {\bf 7}, L273 (1990).
\bibitem{ma}J.S.Chan,K.C.Chan,R.B.Mann, 
preprint WATPHYS TH-94/06
\bibitem{dk}D.A.Konkowski, T.M.Helliwell,
{\em Phys.Rev.D}, (1994) {\bf 50}, 841
\bibitem{pbm} P.R.Brady and 
C.M.Chambers, preprint NCL94-TP13
\bibitem{ori}A.Ori, {\em Phys.Rev.Letters}, {\bf 67}, 1991, 789
\bibitem{ck}M.Carmeli and M.Kaye, {\em Annals of Physics}, {\bf 103},
 97, (1977)
\bibitem{wb}W.Israel and C.Barrab\'{e}s, {\em Physical Review D}, {\bf 43},
1129, (1991)
\bibitem{ab}A.Bonanno, S.Droz, W.Israel and S.M.Morsink, 
{\em Phys.Rev.D}, (1994) {\bf 50}, 7372
\bibitem{ab1}A.Bonanno, S.Droz, W.Israel and S.M.Morsink, 
{\em Proc.R.Soc.Lond. A}, to appear
\end{thebibliography}
\end{document}